\documentclass[runningheads]{cl2emult}

\usepackage{makeidx}  
\usepackage{graphicx} 
\usepackage{subeqnar} 
\usepackage{multicol} 
\usepackage{cropmark} 
\usepackage{eso}      
\makeindex            

\newcommand{\la}{\stackrel{<}{_{\sim}}} 

\begin{document}
%
\title*{Evolutionary models of zero metallicity stars}
\toctitle{Evolutionary models of zero metallicity stars}
%
%
\titlerunning{Evolutionary models of zero metallicity stars}
%
\author{Paola Marigo
\and Cesare Chiosi
\and L\'eo Girardi
\and Tiziana Sarrubbi}
\authorrunning{Paola Marigo et al.: Evolutionary models of zero metallicity stars}
%
%
\institute{Department of Astronomy, University of Padova,
Vicolo dell'Osservatorio 5, 35122 Padova, Italy}

\maketitle              

\begin{abstract}
We present new evolutionary models for zero-metallicity stars,
covering a large range of initial masses (from 0.8 to 100 $M_{\odot}$).
Models are computed with overshooting from stellar cores and 
convective envelopes, and assuming 
mass-loss from the most massive stars. We discuss the main
evolutionary features of these stars,
and provide
estimates of the amount of newly-synthesized elements dredged-up to
the stellar surface, and possibly lost by stellar winds from the most
massive stars.
Full details of these models will be given in Marigo et al. 
(2000, in preparation).
\end{abstract}

\section{Model prescriptions}
In our computations we adopt
an initial helium content $Y=0.23$. We consider 
the complete sets of reactions 
for the pp-chain and CNO tri-cycle, and the most 
important alpha-capture reactions for heavy elements up to Mg.
Nuclear rates are taken from Caughlan \& Fowler (1988).
The nuclear network is implicitly solved for all the considered
H- and He-burning reactions, and without any additional assumption
for nuclear equilibria.
Overshooting from stellar cores is applied according to the 
Bressan et al.\ (1981) formalism.
Mass-loss from massive stars ($M \ge 10 M_{\odot}$)
is described  according to Chiosi (1981). 
Stellar evolution is  calculated throughout the nuclear H-burning
phase up to the He-flash at the RGB tip for low-mass models, and 
throughout nuclear H- and He-burning phases up to  
the beginning of the TP-AGB phase and carbon ignition  
for intermediate- and high-mass models, respectively.
\section{Evolutionary features}
\begin{figure}
	\centering
\resizebox{0.66\textwidth}{!}{\includegraphics{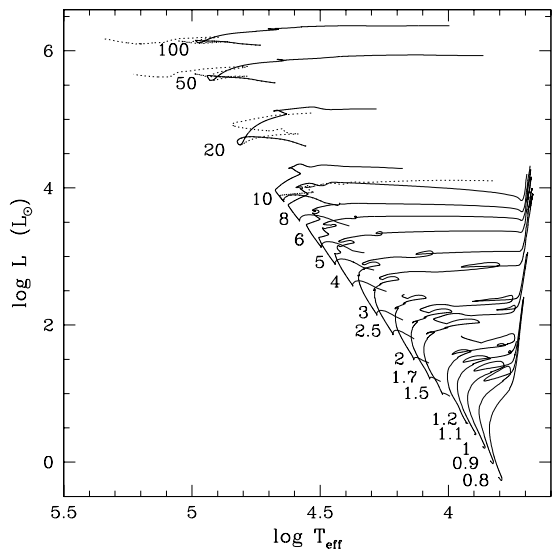}}
\hfill
\parbox[b]{0.29\textwidth}{
	\caption{HR diagram for the complete set of $Z=0$ evolutionary
tracks, from 0.8 to 100 $M_{\odot}$. 
Solid line: evolution at constant mass;  
dotted line: evolution with mass-loss}
	\label{fig_hrd}}
	\end{figure}

Figure~\ref{fig_hrd} presents the evolutionary tracks in the
H-R diagram.  The appearance of short-lived loops
is caused by the 
ignition of the 3-$\alpha$ reaction during core 
and/or shell H-burning, leading to the first 
activation of the CNO-cycle.
In the case of  
low-mass stars ($0.9 M_{\odot} \la M \la 1.2 M_{\odot}$) 
a loop develops  near the end of central H-burning ($X_{\rm c} \sim 0.01$),
whereas for more massive stars ($2.5 M_{\odot} \la M \la 6 M_{\odot}$)
a similar feature  
also occurs at the formation of the He-shell. 

As far as the critical stellar masses for non-degenerate nuclear
ignition are concerned, we find that $M_{\rm Hef} \sim 1.1 M_{\odot}$ 
corresponds to the minimum initial mass for a star to avoid the He-flash, 
and $M_{\rm up} \sim 6.0 - 7.0 M_{\odot}$ the minimum initial mass for 
a star to avoid carbon deflagration.

\section{Surface chemical changes}

The first dredge-up is practically absent in all models. Only for stars with
$M <1.0 M_{\odot}$ a very small amount (less than 0.003 $M_{\odot}$) 
of He is dredged-up
to the surface.
The second dredge-up is experienced by stars with  
$1.5 M_{\odot} < M < 8 M_{\odot}$, being quite
efficient in models with 
$M \ge 2.5  M_{\odot}$. 
It turns out that 
the surface composition is enriched 
almost only in He (reaching $Y = 0.25$ -- $0.37$ at increasing stellar mass) 
and negligibly in CNO elements ($10^{-16}$ -- $10^{-8}$ 
in mass fraction). 

Massive stars, with $10 M_{\odot} < M < 100 M_{\odot}$, do not
show any trace of surface chemical pollution due to dredge-up episodes.
Only models calculated with mass-loss may expose nuclearly processed material
to the surface. 
With the adopted prescription for mass-loss,
stellar winds are never able to strip off H-exhausted layers in models
with $M \le 20 M_{\odot}$. In these cases,
at most, the H-burning shell is eaten up, with consequent surface enrichment
in He and small amounts of CNO.
For more massive models with $M \ge 50 M_{\odot}$, wind stripping
is able to reach the CO-enriched region left by convective He-burning.
In these cases, the corresponding yields of C and O are considerable
($\sim 0.1$ -- $1 M_{\odot}$).

\clearpage
\addcontentsline{toc}{section}{Index}
\flushbottom
\printindex

\end{document}